# Lomics: Generation of Pathways and Gene Sets using Large Language Models for Transcriptomic Analysis

[†]Chun-Ka WONG, MBBS[1]; Ali CHOO, MPhil[1]; Eugene C. C. CHENG, MBBS[1]; Wing-Chun SAN, BSS[1]; Kelvin Chak-Kong CHENG, MBBS[1]; Yee-Man LAU, PhD; Minqing LIN, MBBS MPH; Fei LI, MBBS; Wei-Hao LIANG, MBBS; Song-Yan LIAO, MBBS PhD; Kwong-Man NG, PhD; Ivan Fan-Ngai HUNG, MD; Hung-Fat TSE, MBBS MD PhD[1]; [†]Jason Wing-Hon WONG, PhD[2].

[1]Cardiology Division, Department of Medicine, School of Clinical Medicine, Li Ka Shing Faculty of Medicine, The University of Hong Kong, Hong Kong SAR, China
[2]School of Biomedical Sciences, Li Ka Shing Faculty of Medicine, The University of Hong Kong, Hong Kong SAR, China
[†]These authors are co-corresponding authors

**Short title:** Lomics: Large Language Models for Omics Studies
**Word count:** 1437

**Address for correspondence:**
Chun-Ka Wong, MBBS
Clinical Assistant Professor
Department of Medicine, Li Ka Shing Faculty of Medicine,
The University of Hong Kong, Hong Kong SAR, China
Tel: +852-22553597; Fax: +852-28186304
Email: wongeck@hku.hk

Jason Wing-Hon WONG, PhD
Professor
School of Biomedical Sciences, Li Ka Shing Faculty of Medicine,
The University of Hong Kong, Hong Kong SAR, China
Tel: +852- 39179187; Fax: +852- 28170857
Email: jwhwong@hku.hk

# ABSTRACT

Interrogation of biological pathways is an integral part of omics data analysis. Large language models (LLMs) enable the generation of custom pathways and gene sets tailored to specific scientific questions. These targeted sets are significantly smaller than traditional pathway enrichment analysis libraries, reducing multiple hypothesis testing and potentially enhancing statistical power. Lomics (Large Language Models for Omics Studies) v1.0 is a python-based bioinformatics toolkit that streamlines the generation of pathways and gene sets for transcriptomic analysis. It operates in three steps: 1) deriving relevant pathways based on the researcher's scientific question, 2) generating valid gene sets for each pathway, and 3) outputting the results as .GMX files. Lomics also provides explanations for pathway selections. Consistency and accuracy are ensured through iterative processes, JSON format validation, and HUGO Gene Nomenclature Committee (HGNC) gene symbol verification. Lomics serves as a foundation for integrating LLMs into omics research, potentially improving the specificity and efficiency of pathway analysis.

# ABBREVIATION

**HGNC**  HUGO Gene Nomenclature Committee
**KEGG**  Kyoto Encyclopedia of Genes and Genomes
**LLM**  Large Language Model

# APPLICATION NOTES

## 1. Introduction

Broadly speaking there are two approaches with which researchers can derive insights from high-throughput omics data generated from biological experiments, including hypothesis-driven approach and unbiased approach.[1] In hypothesis-driven approach, researchers list out a relatively small number of relevant pathways and gene sets, typically at the scale of tens of pathways, based on their knowledge and to perform targeted analysis of the obtained transcription data. In unbiased approach, researchers make no presumption about target pathways and perform pathway enrichment studies large number of pathways, typically thousands, using standard libraries such as Kyoto Encyclopedia of Genes and Genomes (KEGG).[2] Hypothesis-driven approach can be time-consuming and is limited by the breadth of knowledge and depth of insights that an individual researcher may possess. On the other hand, the cost of performing unbiased approach is the need to adjust for statistical significance owing to multiple hypothesis testing.[3-4] In this study, we explore a third approach that sits between the two classical methods for analyzing transcription data: an artificial intelligence-facilitated hypothesis-driven approach.

Recently, there have been rapid advances in the development of foundation large language models (LLMs).[5-6] Large volume of biological knowledge is stored in the weights of these LLMs. Previous studies have demonstrated the possibility of utilizing LLMs to derive innovative insights that are not present in the original training sets.[7] Using foundation LLMs, we developed the Lomics (Large Language Models for Omics Studies), which is a tool for researchers to quickly generate relevant pathways and valid gene sets for any given scientific question using foundation LLMs.

## 2. Implementation
### 2.1. Lomics overview

Lomics v1.0 is python-based bioinformatics pipeline that consists of three key steps. First, it derives a list of relevant pathways for the scientific question submitted by researchers. Second, it generates valid gene sets for each pathway. Third, it outputs the generated pathways and gene sets as .GMX files. Lomics also outputs an explanation of why each pathway has been generated.

## 2.2. LLM backbone

Lomics employs a flexible architecture that can integrate with various state-of-the-art foundation Large Language Models (LLMs). This design allows researchers to leverage the latest advancements in LLM technology without being constrained to a specific model. The system utilizes LiteLLM v1.40.29 to facilitate application program interface (API) calls, ensuring seamless communication with diverse LLM services and cloud platforms. Asyncio v3.4.3 is used to enable parallel processing.

## 2.3. Pathway generation

Lomics outputs a list of pathways for enrichment analysis for any given scientific question (i.e. a prompt) that researchers specify. Researchers can adjust the number of pathways to be generated, with the default value being 40. To address inconsistencies in current generation LLMs in generating the exact number of pathways as instructed, Lomics explicitly instructs LLMs to output a JSON object with keys "pathway 1", "pathway 2", ..., "pathway P", where P is the specified number of pathways. This structured output format improves reliability compared to simply requesting the LLM to return P pathways. Pydantic v2.7.4 is used to validate that the LLM output adheres to the prespecified format, ensuring consistency and accuracy.

LLMs may produce varied outputs for the same prompt due to the use of random seeds in result generation. To enhance the reproducibility of pathway generation, Lomics runs the same prompt multiple times, with a default of 10 iterations. Temperature is set as 0. The most frequently proposed pathways across these iterations are then presented to researchers.

## 2.4. Gene set generation

Lomics uses LLMs to output gene sets for each pathway obtained from the previous step. The size of gene sets can be specified by researchers, with a default value of 40 genes per pathway. Similar to the previous step, Lomics explicitly instructs LLMs to output a JSON object with keys "gene 1", "gene 2", ..., "gene G", where G is the specified number of genes, to ensure exact number of genes are returned, before passing results to Pydantic for validation. To ensure validity of the proposed genes,

LLMs are instructed to output HUGO Gene Nomenclature Committee (HGNC) gene symbols. HGNC database is used to ensure the validity of the derived gene symbols.

To enhance gene set generation consistency, Lomics runs the same prompt multiple times, with a default of 5 iterations. Temperature is set as 0. The most frequently proposed genes across these iterations are then presented to researchers.

**2.5. Availability**

Codes of Lomics v1.0 are available on GitHub (https://www.github.com/wongchunka/lomics). By default, LLM API calls are mediated via LiteLLM with Replicate (Replicate, USA) and LLama-3 70B Instruct model (Meta, USA).[5] API key for making LLM calls have be inserted in setting.py before use. Choice of LLM and cloud provider can be modified following instructions of LiteLLM:

```
os.environ["REPLICATE_API_KEY"] = "YOUR_API_KEY"
var_llm = "replicate/meta/meta-llama-3-70b-instruct"
```

Other Lomics settings, such as number of pathways and genes to be generated, as well as number of replicates, can similarly be modified in setting.py:

```
var_num_pathway = 40
var_iterate_pathway = 5
var_num_gene = 40
var_iterate_gene = 2
```

Lomics can be run using python v3.9.5 with the command line interface (CLI) code listed below. A wide variety of scientific questions can be used, for instance "mechanism of SARS-CoV-2 induced myocardial injury" and "effect of SARS-CoV-2 on the endothelial cells".

```
python run.py --question "scientific question" --outputname "output file name" --outputdir "output file directory"
```

Alternatively, a freely accessible web application is also available (https://www.lomics.ai). **(Figure 1A)** The web application version currently utilizes Replicate to serve LLama-3 70B Instruct as the backbone LLM.[5] It returns top 40 pathways for each scientific question after iterating for 5 times. It returns top 40 genes for each pathway after iterating for 2 times. Temperature is set as 0 for all steps.

## 3. Use cases

### 3.1. Pathway enrichment analysis for transcriptomic studies

The generated pathways and gene sets summarized in a .GMX file can be imported for pathway enrichment analysis in RNA sequencing experiments, using tools such as clusterProfiler in R and GSEApy in Python.[8-9] Comparative studies between Lomics and conventional unbiased pathway enrichment analysis using standard libraries like KEGG can be performed to derive biological insights.[2] The number of pathways generated by Lomics (40 pathways) is significantly smaller than standard libraries such as KEGG (571 pathways). As a result, the need for correcting statistical significance for multiple hypothesis testing using methods such as the Benjamini-Hochberg procedure is much reduced. [3,10] Furthermore, Lomics outputs an explanation for each suggested pathway, enabling researchers to critically evaluate the validity and potential of the proposed mechanistic association with the scientific question of interest.

### 3.2. Augmenting hypothesis-driven studies

Lomics output may also enhance hypothesis-driven research. Traditionally, when researchers conduct targeted analysis of biological specimens based on hypotheses- such as quantitative polymerase chain reaction (qPCR) of specific pathways and genes- they are constrained by their individual knowledge and insights. Lomics may help overcome this limitation by rapidly generating many plausible pathways to investigate, thus expanding the scope beyond personal expertise.

### 3.3. Application of Lomics for studying mRNA vaccine associated myocarditis

We applied Lomics v1.0 to bulk RNA sequencing data generated from experiments that investigated the mechanism of mRNA vaccine-associated myocarditis. There was empirical evidence that higher mRNA vaccine dosage was associated with a higher risk of developing myocarditis in children and young adolescence. [11] Induced pluripotent stem cell-derived cardiomyocytes (iPSC-CMs) from the IMR-90 cell line were cultured in high dose (0.4 microgram/ml) and low dose (0.0004 microgram/ml) of BNT162b2 mRNA vaccine (Pfizer, USA). Cell lysates from treatment and control groups were collected on day 1 for RNA extraction.

Pathway enrichment analysis was performed using pathways from both Lomics and KEGG. Lomics v1.0 was used to generate pathways and gene sets with the question, "Study the mechanism of mRNA vaccine induced myocarditis using cardiomyocytes". LLama-3 70B Instruct was used as the base model. Other parameters used include number of pathways of 40, number of genes per pathway of 40, iterations for pathway generation 10, iteration for gene set generation 5, and temperature of 0. GSEApy v1.1.1 with posterior probability of differential expression (PPDE) > 0.95 was used in both approaches.[9]

Comparison of results derived from Lomics and KEGG is shown in Figure 1D. Standard approach using KEGG library only yielded 1 pathway ("steroid biosynthesis") that was significantly enriched (adjusted p=0.0002*). On the other hand, 3 out of the 40 pathways generated by Lomics specific for the scientific question yielded positive results, including cardiac fibrosis, cytosolic DNA-sensing pathway, and endoplasmic reticulum stress (adjusted p=0.01*). **(Figure 1B)** This case study demonstrates the potential of Lomics to serve as a complementary approach for generating gene sets for pathway enrichment studies.

## 4. Conclusion

Lomics is a novel bioinformatics tool that allows researchers to generate pathways and gene sets using large language models for transcriptomic analysis.

**Contribution statement:**

CKW and JWHW contributed to the conception and design of the study. CKW, AC, ECCC, WCS, KCCC, HFT and JWHW developed the Lomics system. YML, ML, FL, WHL, SYL, KMN, HFT, and IFNH contributed to in-vitro studies. CKW, AC, ECCC, and JWHW wrote first draft of the manuscript. WCS, KCCC, YML, ML, FL, WHL, SYL, KMN, IFNH and HFT revised the manuscript critically for important intellectual content. All authors have read and approved the final version of the manuscript to be published.

**Disclosure:**

The authors report no conflict of interest.

**Funding:**

This study was partly supported by the Li Shu Fan Medical Fellowship for Internal Medicine by Li Shu Fan Medical Foundation and Li Ka Shing Faculty of Medicine, The University of Hong Kong; and Seed Fund for Basic Research for New Staff, University of Hong Kong.

**Acknowledgement:**

None.

**Data availability statement:**

Codes of Lomics v1.0 are available on GitHub (https://www.github.com/wongchunka/lomics). A freely accessible web application is also available (https://www.lomics.ai).

# FIGURE LEGEND

**Figure 1. Lomics: Large Language Models for Omics Studies.** (A) Example of Lomics web application showing list of proposed pathways, gene sets, and explanations. (B) Comparison of pathway enrichment analysis with GSEApy using KEGG and Lomics generated gene sets. Total RNA from hiPSC-CMs cultured with high versus dose BNT162b2 mRNA vaccine in triplicate was analyzed.

# REFERNCES

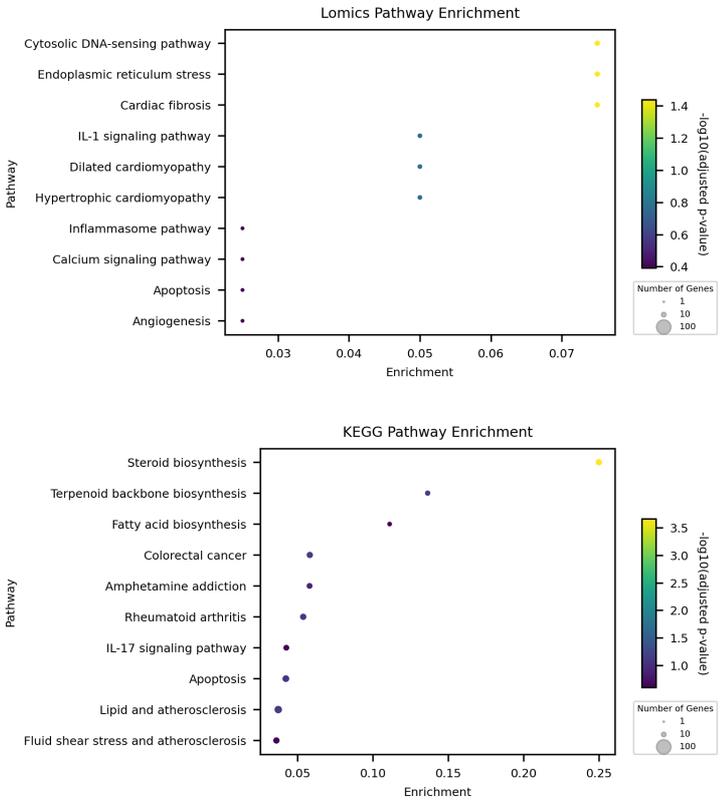

Figure 1